\documentclass[twocolumn, superscriptaddress, aps, prl, reprint]{revtex4-2}
\usepackage[english]{babel}
\usepackage[utf8]{inputenc}
\usepackage[colorinlistoftodos, color=green!40, prependcaption]{todonotes}
\usepackage{multirow}
\usepackage{url}

\usepackage{siunitx}

\usepackage[pdftex, pdftitle={Article}, pdfauthor={Author}]{hyperref} 

\usepackage{graphicx}
\usepackage{dcolumn}
\usepackage{bm}

\begin{document}

\title{Infrared photon-number-resolving imager using a Skipper-CCD} 

\author{Q. Pears Stefano}
\email{quimeyps@gmail.com}
\affiliation{Universidad de Buenos Aires, Facultad de Ciencias Exactas y Naturales, Departamento de Física, Pabellón I, Ciudad Universitaria (1428), Buenos Aires, Argentina \\}
\affiliation{CONICET - Universidad de Buenos Aires, Buenos Aires, Argentina.\\}

\author{A. G. Magnoni}%
\email{magnoni.agustina@gmail.com}
\affiliation{Laboratorio de Óptica Cuántica, DEILAP, UNIDEF (CITEDEF-CONICET), Buenos Aires, Argentina}%
\affiliation{Universidad de Buenos Aires, Facultad de Ciencias Exactas y Naturales, Departamento de Física, Pabellón I, Ciudad Universitaria (1428), Buenos Aires, Argentina \\}

\author{J. Estrada}
\affiliation{Fermi National Accelerator Laboratory, Batavia IL, United States}%

\author{C. Iemmi}%
\affiliation{Universidad de Buenos Aires, Facultad de Ciencias Exactas y Naturales, Departamento de Física, Pabellón I, Ciudad Universitaria (1428), Buenos Aires, Argentina \\}
\affiliation{CONICET - Universidad de Buenos Aires, Buenos Aires, Argentina.\\}

\author{D. Rodrigues}
\affiliation{Universidad de Buenos Aires, Facultad de Ciencias Exactas y Naturales, Departamento de Física, Pabellón I, Ciudad Universitaria (1428), Buenos Aires, Argentina \\}
\affiliation{CONICET - Universidad de Buenos Aires, Instituto de Física de Buenos Aires (IFIBA), Buenos Aires, Argentina.\\}

\author{J. Tiffenberg}
\affiliation{Fermi National Accelerator Laboratory, Batavia IL, United States}%

\date{\today}

\begin{abstract}

Imaging in a broad light-intensity regime with a high signal-to-noise ratio is a key capability in fields as diverse as Quantum Metrology and Astronomy. Achieving high signal-to-noise ratios in quantum imaging leads to surpassing the classical limit in parameter estimation. In astronomical detection, the search for habitable exoplanets demands imaging in the infrared its atmospheres looking for biosignatures.
These optical applications are hampered by detection noise, which critically limits their potential, and thus demands photon-number and spatial resolution detectors.
Here we report an imaging device in the infrared wavelength range able to arbitrarily reduce the readout noise. We built a Measured Exposure Skipper-CCD Sensor Instrument equipped with a thick back-illuminated sensor, with photon-number-resolving capability in a wide dynamic range, spatial resolution, high quantum efficiency in the near-infrared and ultra-low dark counts. This device allows us to image objects in a broad range of intensities within the same frame and, by reducing the readout noise to less than 0.2e$^-$, to distinguish even those shapes with less than two photons per pixel, unveiling what was previously hidden in the noise. These results pave the way for building high-standard infrared imagers based on Skipper-CCDs.
\end{abstract}


\maketitle


\section{Introduction}\label{sec:intro}

The search for a spatially resolved sensor with photon-number resolving capability in a wide dynamic range for optical and near-infrared wavelengths had been a challenge shared by both Quantum Optics and Astronomy. Among the desirable characteristics of such detectors, a high quantum efficiency (QE) and low dark current (DC) are the most relevant \cite{2022SPIE12180E..65R,Crill2019, Moreau2019, berchera2019quantum}.
Thus, the ultra-low readout noise Skipper-CCD~\cite{Tiffenberg2017}, until now only used as a particle detector \cite{SENSEI2020, DAMICM2020, OSCURA2022, CONNIE2021}, arises as an excellent candidate to fulfil all these requirements. 
The Skipper-CCD photon-number resolving capability is achieved by eliminating the low-frequency readout noise (around 2e$^-$ in a conventional CCD) by multiple, non-destructive measurements (samples) of the charge in each pixel. With this method, photon counting is possible over a wide dynamic range as the noise is independent of the charge deposited in the pixel~\cite{Rodrigues2021}.  
For a readout noise of $\sim$0.2e$^{-}$ (attained after 256 samples) the probability of misclassification is lower than 1\% and can be further reduced just by increasing the number of charge samples. 

A very attractive feature of this thick (200~$\mu$m) fully depleted sensor, is that they offer a QE above 80\% at 800~nm~\cite{Drlica2020}, and it is expected to hold this value until 950~nm based on the performance of similar CCDs~\cite{Bebek2017}. 
This technology also presents the lowest DC in literature with demonstrated $\sim$10$^{-4}$ e$^{-}$/pix/day \cite{SENSEI2022}. For many applications in quantum optics and astronomy, having all these features available in a spatially resolved sensor with only 15~$\mu$m of pixel size results in a disruptive technology~\cite{ Magnoni2021}.

In Quantum Optics, especially in metrology, detecting 800~nm photons with a high QE, and essentially no noise, enhances the performance of several ongoing and proposed applications such as Quantum Microscopy \cite{Samantaray2017} and Quantum Illumination~\cite{Gregory2020} among others \cite{Brida2010, Moreau2019}. At the same time, in Astronomy, all these features, along with a high radiation tolerance, stability, linear response, and a full-well capacity above 33000 electrons~\cite{Drlica2020} make the Skipper-CCDs especially suited for the search of signatures of life in Earth-like exoplanets~\cite{Marrufo2022}.

Competing technologies such as CMOS or EMCCDs recently demonstrated sub-electron noise capability~\cite{Hamamatsu2021}, however, the trade-off between low DC and the thickness required to get relatively high QE for infrared wavelengths, results in an unacceptable performance~\cite{Harding2015}.
Also, the effective readout noise of EMCCD depends on the gain and is restricted to a very short occupancy range.
The current clear advantage of CMOS over CCD is the shorter readout time, which allows for many frames per second. However, ongoing efforts to reduce the Skipper-CCD readout time could make possible a factor of one hundred faster readings compared with the current performance~\cite{Chierchie2020}. Other options for photon-number-resolving sensors proposals, based on multiplexing of Superconducting Nanowires Single-photon and Transition-edge detectors have shorter dynamic ranges and lack spatial resolution~\cite{cheng2022100, Eaton2023}.   

Here we report the use of Skipper-CCD for imaging purposes for the first time. We demonstrated that information hidden within the readout noise can be revealed after performing multiple non-destructive charge measurements. A set of pictures with intensities ranging from 2 to 40 photons per pixel on average were taken at 810~nm. 
In one of them, a text summarising this work appears after reducing the readout noise to 0.25e$^{-}$. 

\section{Experimental Setup}\label{sec:experimental}
\begin{figure}[th]
  \includegraphics[width=.90\linewidth]{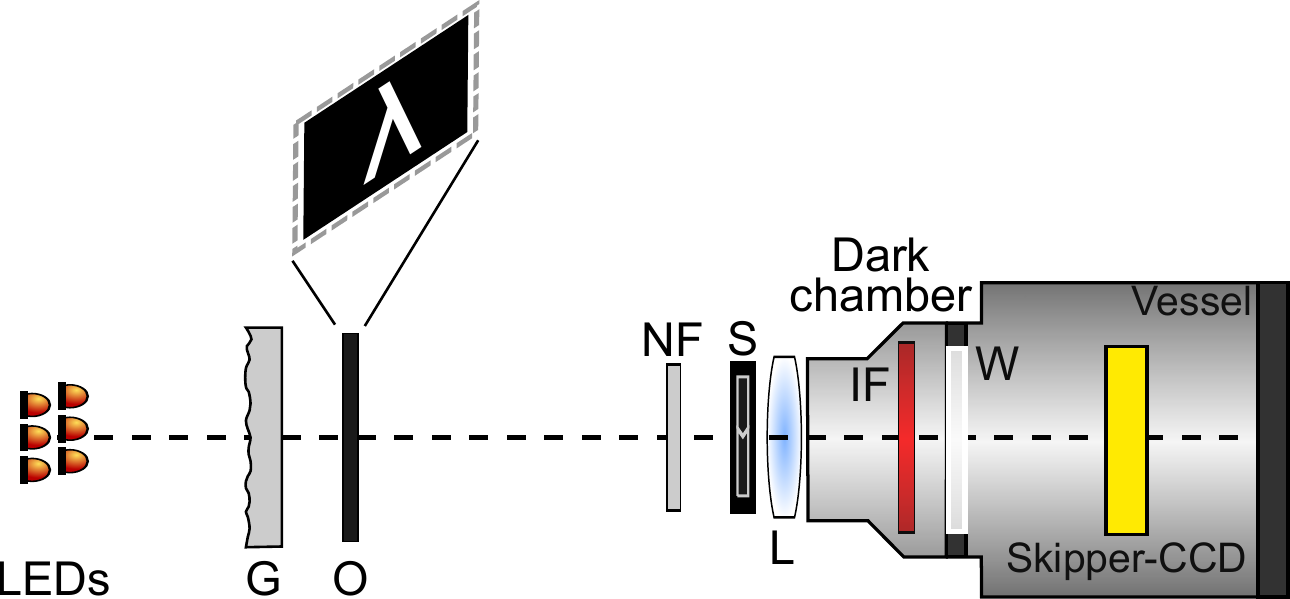}
  \caption{Schematic view of the experimental setup. The light of six infrared LEDs is diffused in a ground glass (\textbf{G}), before illuminating the object (\textbf{O}). An image is formed  on a Skipper-CCD with a convergent lens (\textbf{L}), and light exposure is controlled with a shutter (\textbf{S}). Finally, the light intensity is attenuated to work in a very low-light regime by means of two filters (\textbf{NF} and \textbf{IF}). The image is not to scale.
  \label{fig:setup}}
\end{figure}

Figure \ref{fig:setup} shows a schematic view of the Measured Exposure Skipper-CCD Sensor Instrument setup. The light from an array of six infrared LEDs (central wavelength \SI{865}{\nm}, FWHM \SI{40}{\nm}) impinges onto a ground glass (\textbf{G}) resulting in an approximately uniform non-coherent light source that illuminates a mask that constitutes the object to be imaged.

 The optical system consists of a convergent lens ({\bf L}) of \SI{18}{\cm} focal length and an electronic shutter ({\bf S}) Melles Griot of \SI{25.4}{\mm} diameter, placed in front of the lens, used to control light exposure. Additionally, neutral filters ({\bf NF}) are employed in order to reduce the light intensity. The lens was placed at \SI{21}{\cm} from the Skipper-CCD's active area, and the object ({\bf O}) was placed at approximately \SI{120}{\cm} in front of the lens. This resulted in a lateral magnification of about \num{0.17}. The object consists of a cut-out text (or shape) over an opaque material.
 
Extra precautions were taken to avoid spurious ambient light from entering the sensor. A dark chamber composed of a black-painted metallic holder was firmly attached to the Skipper-CCD vessel. This metallic holder was used to adjust the lens position. Additionally, in front of the fussed silica window {\bf W}, a square interferential filter ({\bf IF}) centered at \SI{810}{\nm} (with an FWHM \SI{10}{\nm}) was placed. This filter allows to pass of about $2\%$ of the light provided by the LEDs, thus reducing the total intensity. Furthermore, the illumination-object system was enclosed in a metallic foil-lined box, with a small aperture in the place of the object mask. All measurements were performed in a dark room.

The Skipper-CCD sensor was designed by the Lawrence Berkeley National Laboratory (LBNL) and fabricated at Teledyne/DALSA using high-resistivity ($>$10k$\Omega$-cm) silicon wafers. It operates in the range of $\SI{135}{K}$ to $\SI{140}{K}$, in vacuum, and has a pixel size of $\SI{15}{\mu\metre}$ by $\SI{15}{\mu\metre}$ with a total active area of $4126\times 886$ pixels. The system can be configured to only perform $N$ charge samples in the region of interest of each image, hence reducing the overall readout time.

\section{Image processing and results}\label{sec:results}

\begin{figure*}[hbtp]
\centering
  \includegraphics[width=1\linewidth]{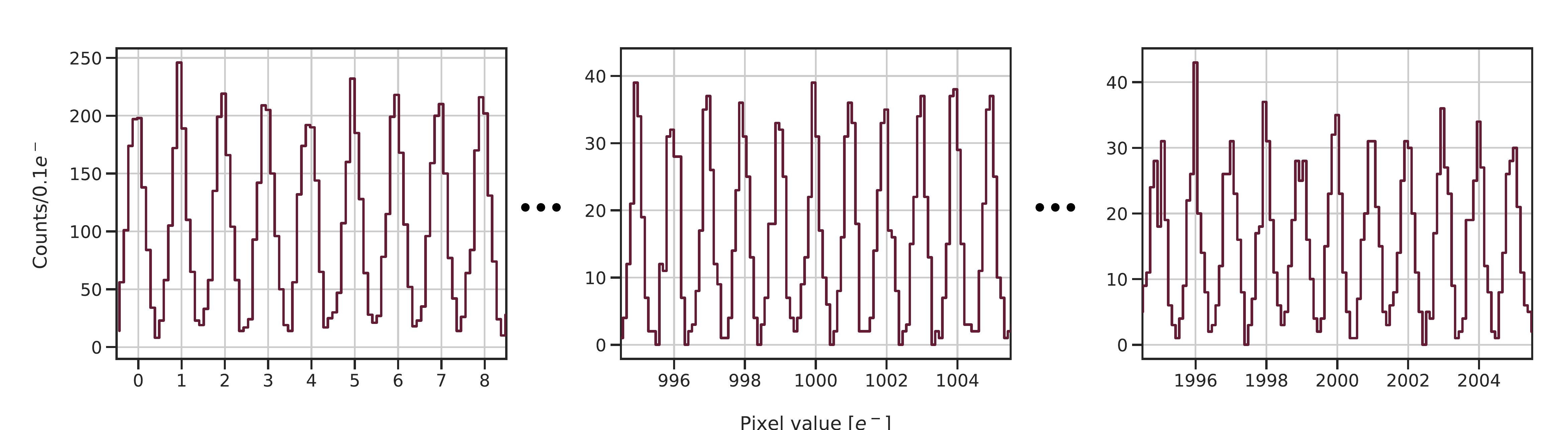}
  \caption{Charge histograms in three different regions of the sensor dynamic range. Very low occupancy (left), around a thousand charges per pixel (middle), and high occupancy (right). The same sub-electron peak resolution is observed in the full range, which was achieved after $N=300$ samples.}
  \label{fig:high_occupancy}
\end{figure*}

Once the sensor has been exposed to light, the reading of the pixel values is performed for the desired amount of charge samples $N$. This produces raw data in Analogical to Digital Units (ADUs). Then, the samples are averaged and the ADU values are converted to electrons using a calibration determined by the separation between the peaks in the recorded charge histogram that correspond to zero and one electron, as described in~\cite{Rodrigues2021}. In the case of high occupancy, this procedure can be performed up to 2000e- as shown in Figure \ref{fig:high_occupancy}.

\begin{figure*}[hbtp]
\centering
  \includegraphics[width=1\linewidth]{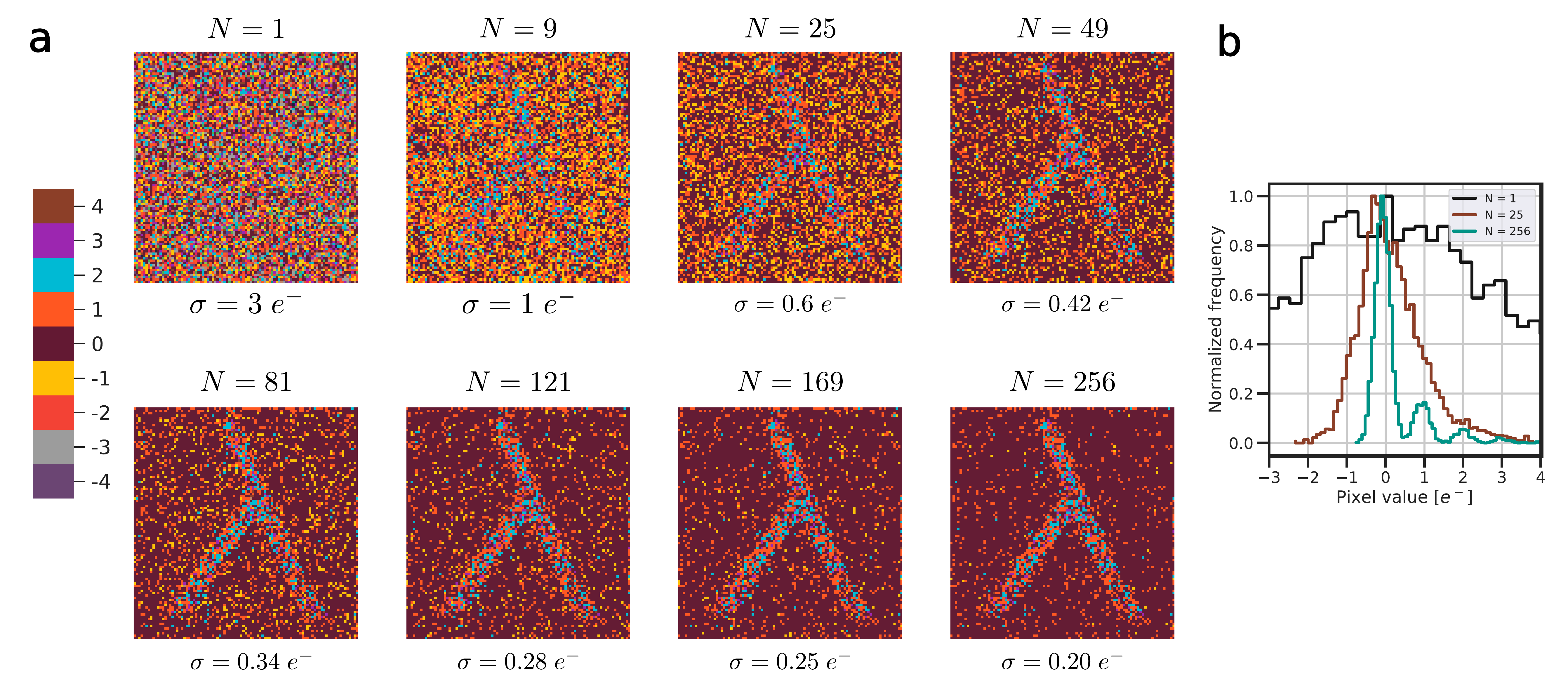}
  \caption{Results for the imaging of the greek letter $\lambda$ with a Skipper-CCD. \textbf{a}: images with different amounts of charge samples ($N$) from 1 to 256. The color map used is discrete in the number of electrons measured. \textbf{b}: measured charge histogram of a region of the image, for $N = \{1, 25, 256\}$.}
  \label{fig:lambda}
\end{figure*}

Figure \ref{fig:lambda} \textbf{a} shows a $98\times 95$ pixels image of the greek letter $\lambda$ with $N$ ranging from 1 to 256. The readout noise ($\sigma$) starts in 3e$^-$ for $N =1$, and scales down with $\sqrt{N}$, even reaching the sub-electron regime. For $N =1$, meaning that only one measurement of the charge in each pixel is taken, the mean number of photons inside the area of $\lambda$ takes lower values (ranges from 0.7 to 1.2) than the readout noise. As a result, the image is hidden and all the information is invisible. However, with increasing $N$, the readout noise decreases from 3e$^-$ to 0.2e$^-$, improving the signal-to-noise ratio and allowing $\lambda$ to appear at plain sight.

Figure~\ref{fig:lambda} \textbf{b} shows the measured charge histogram for three representative values of $N$: 1, 25, and 256. The reduction in the readout noise to deep sub-electron values produces a discrete charge histogram, where the different peaks for integer values of electrons can be resolved independently.       

Additionally, video \ref{vid:ITNWSKP} shows the $280\times 460$ pixels image containing a text summarising this work for different values of $N$ in the form of an animated gif. The mean amount of light per pixel in the letters ranges from 1 to 1.5 photons with a mean background of 0.05 photons. 

\begin{video}[h]
\href{https://drive.google.com/file/d/16Ds6yOtBT8NEvo4VD0kvp-bsCVvHSZqq/view?usp=sharing}{\includegraphics[width=1\linewidth]{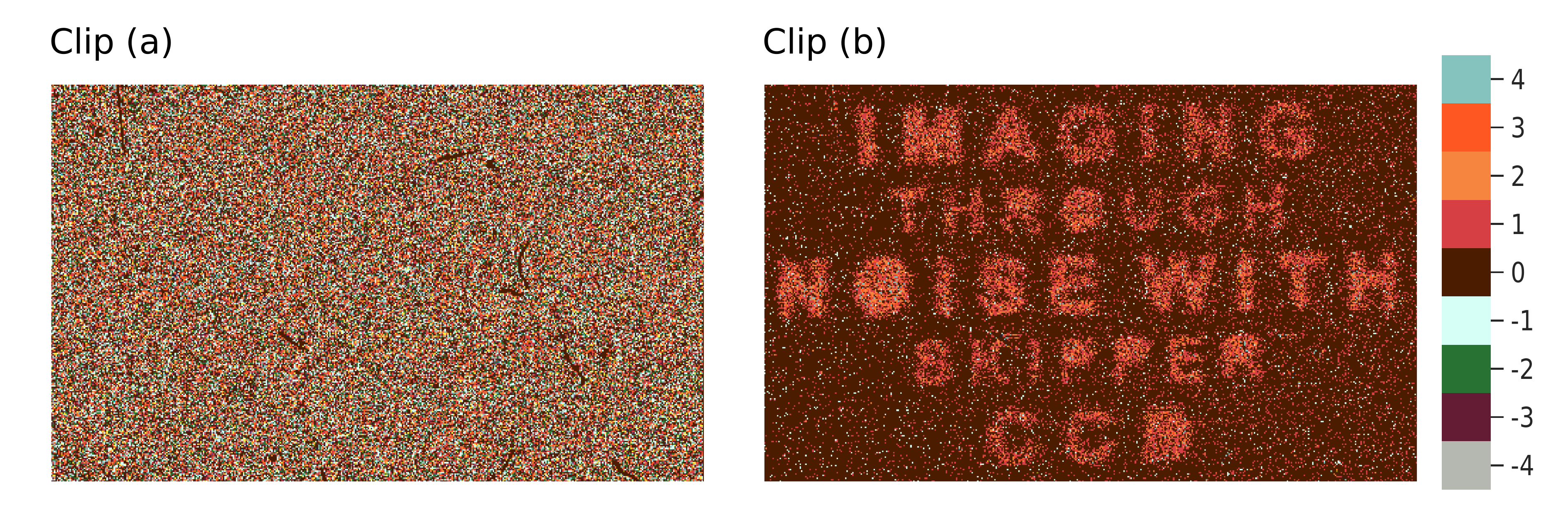}}%
 \setfloatlink{https://drive.google.com/file/d/16Ds6yOtBT8NEvo4VD0kvp-bsCVvHSZqq/view?usp=sharing}%
 \caption{%
  Animated gif of on image describing this work, for different values of $N$. 
  Clip (a): image for $N = 1$, with a readout noise of $\sigma = 3$ e$^-$.
  Clip (b): image for $N = 150$, with a readout noise of $\sigma = 0.24$ e$^-$.}%
  \label{vid:ITNWSKP}
\end{video}

Clip (a) corresponds to $N=1$, where the readout noise masks out the shape of the text, and clip (b) shows the same image with $N = 150$ samples of the charge of each pixel, revealing the hidden image. It is evident in the figure the increase in the signal-to-noise ratio that can be obtained with the use of a Skipper-CCD, allowing access to a low-intensity regime unable to be explored with conventional CCD cameras. 

It is interesting to note that the capability to resolve photon number in a wide dynamic range of the Skipper-CCD opens up the regime of very few (but not only one) photons for imaging purposes. To show this specific feature of the detector, video \ref{vid:estrellas} shows an image of three stars: two of them are more intense (mean photon number of 18) and the third one has a mean photon number of approximately 1 that makes it impossible to distinguish it from the readout noise in the $N=1$ case.

\begin{video}[h]
\href{https://drive.google.com/file/d/1xOqcujjXMbn9ZJyi1Uocd3ebzB_zaEBH/view?usp=sharing}{\includegraphics[width=1\linewidth]{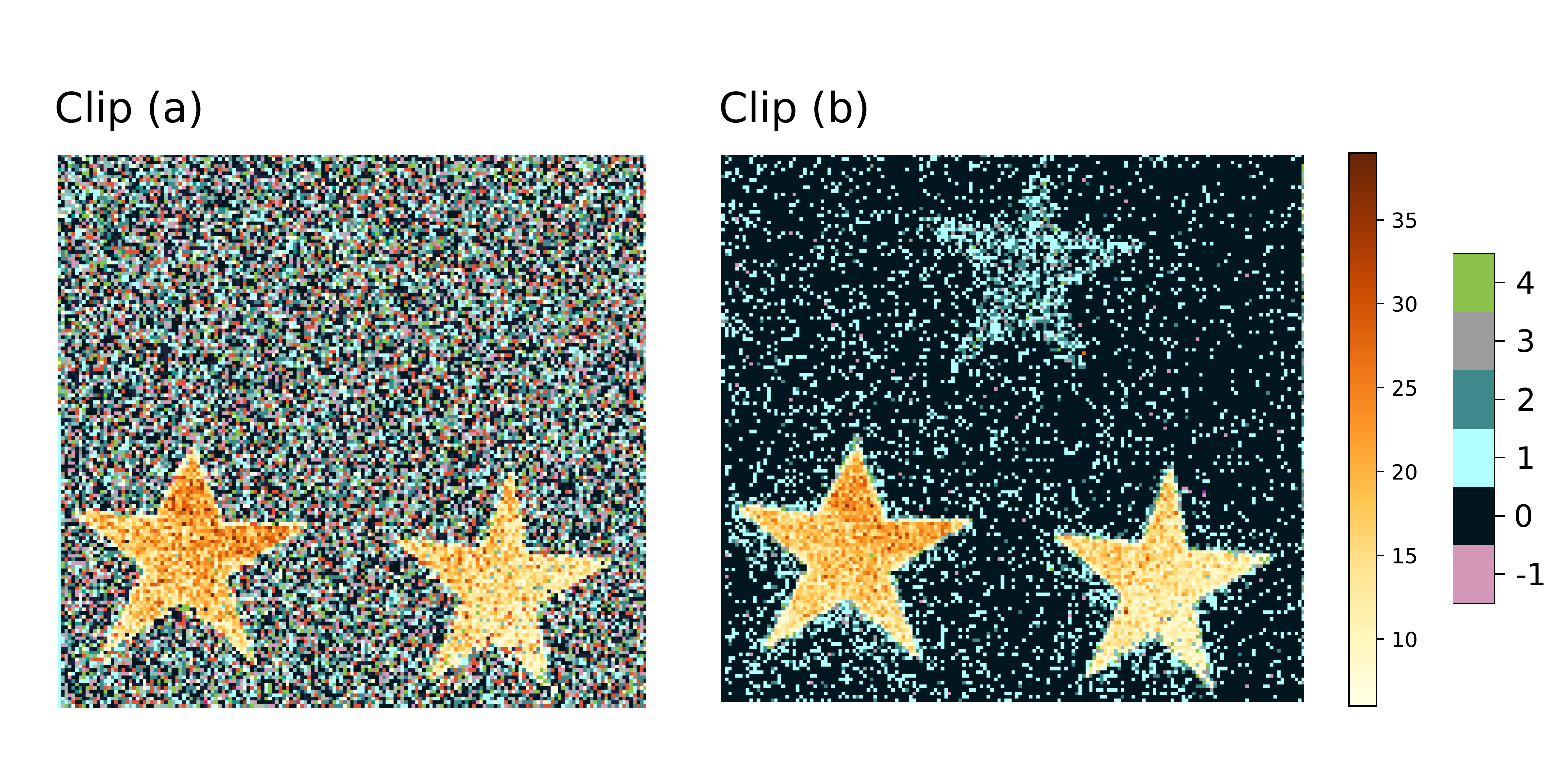}}%
 \setfloatlink{https://drive.google.com/file/d/1xOqcujjXMbn9ZJyi1Uocd3ebzB_zaEBH/view?usp=sharing}%
 \caption{%
  Animated gif of the image of three stars with different mean photon number, for different values of $N$. 
  Clip (a): image for $N = 1$, with a readout noise of $\sigma = 3$ e$^-$.
  Clip (b): image for $N = 256$, with a readout noise of $\sigma = 0.20$ e$^-$.}%
  \label{vid:estrellas}
\end{video}

While the two more intense stars can be seen and could be well resolved by a conventional CCD ($N=1$), only when reducing the readout noise the third star appears. This image demonstrates the capability of the Skipper-CCD to detect objects in different intensity regimes, enlarging the dynamic range in which images with a high signal-to-noise ratio can be obtained.  

\section{Conclusions}\label{sec:conclusions}
 
There are two sources of noise beyond the object when taking images: those that come from the source (shot noise) and those that originate from the sensor (readout noise). The latter can be eliminated by using a Skipper-CCD sensor which, by contrast with other technologies, holds this feature in an unprecedented wide dynamic range from zero to thousand photons per pixel. In this work, objects producing fewer than two photons per pixel on average were imaged after removing readout noise. Therefore, we demonstrate the ability to image in an ultra-low photon regime reducing drastically the second contribution. Ongoing efforts are focused on improving this performance by removing the source contribution using correlated photons. This first through-noise imaging with Skipper-CCD also demonstrates its unique capability for near-infrared imaging for astronomical purposes.

       
\section*{Acknowledgments}

This work was supported by Fermilab under DOE Contract No.\ DE-AC02-07CH11359. 
The CCD development work was supported in part by the Director, Office of Science, of the DOE under No.~DE-AC02-05CH11231.  DR acknowledges the support of Agencia Nacional de Promoción de la Investigación, el Desarrollo Tecnológico y la Innovación through grant PICT 2018-02153. CI and QPS acknowledges the support of the Secretaría de Ciencia y Técnica, Universidad de Buenos Aires (20020170100564BA), of the Consejo Nacional de Investigaciones Científicas y Técnicas (PIP 2330), and of the Ministerio de Ciencia Tecnologia e Innovacion/Agencia Nacional de Promocion Cientifica y Tecnologica (PICT-2020-SERIEA-02031). QPS holds a CONICET Fellowship.


\bibliography{apssamp}

\end{document}